\documentclass{PoS}
\usepackage[FIGBOTCAP,TABBOTCAP,tight]{subfigure}
\usepackage{tikz}
\usepackage{graphicx}
\usepackage{grffile}
\usepackage{overpic}
\usepackage[utf8]{inputenc}
\usepackage{amsmath}
\graphicspath{{./fig/}}

\title{
On the $\beta$- and quark mass dependence of the nuclear transition in the strong coupling regime
}

\ShortTitle{$\beta$ and quark mass dependence of nuclear transition}

\author{\speaker{Jangho Kim}\\
        Institut f{\"u}r Theoretische Physik, Goethe-Universit{\"a}t Frankfurt am Main, Max-von-Laue-Str. 1, 60438 Frankfurt am Main, Germany\\
        E-mail: \email{jkim@th.physik.uni-frankfurt.de}}

\author{Owe Philipsen\\
        Institut f{\"u}r Theoretische Physik, Goethe-Universit{\"a}t Frankfurt am Main, Max-von-Laue-Str. 1, 60438 Frankfurt am Main, Germany\\
        E-mail: \email{philipsen@th.physik.uni-frankfurt.de}}
\author{Wolfgang Unger\\
        Fakult{\"a}t f{\"u}r Physik, Universit{\"a}t Bielefeld, Universit{\"a}tstasse 25, D33619 Bielefeld, Germany\\
        E-mail: \email{wunger@physik.uni-bielefeld.de}}

\abstract{
Lattice QCD in a dual formulation with staggered fermions is well established in
the strong coupling limit and allows to perform Monte Carlo simulations at finite
baryon chemical potential.
We have recently addressed the dependence of the nuclear critical end point as a function of the quark mass $am_q$, and separately as a function of the lattice gauge coupling $\beta$ in the chiral limit.
Here we proceed to determine the dependence of the nuclear transition on both, $am_q$ and $\beta$, on isotropic lattices and attempt to pinpoint the critical end point for various $\beta$ where the sign problem is still manageable.
}

\FullConference{37th International Symposium on Lattice Field Theory - Lattice2019\\
		16-22 June 2019\\
		Wuhan, China}

\begin{document}

\section{Introduction}
\begin{figure}
\center
\includegraphics[width=0.6\textwidth]{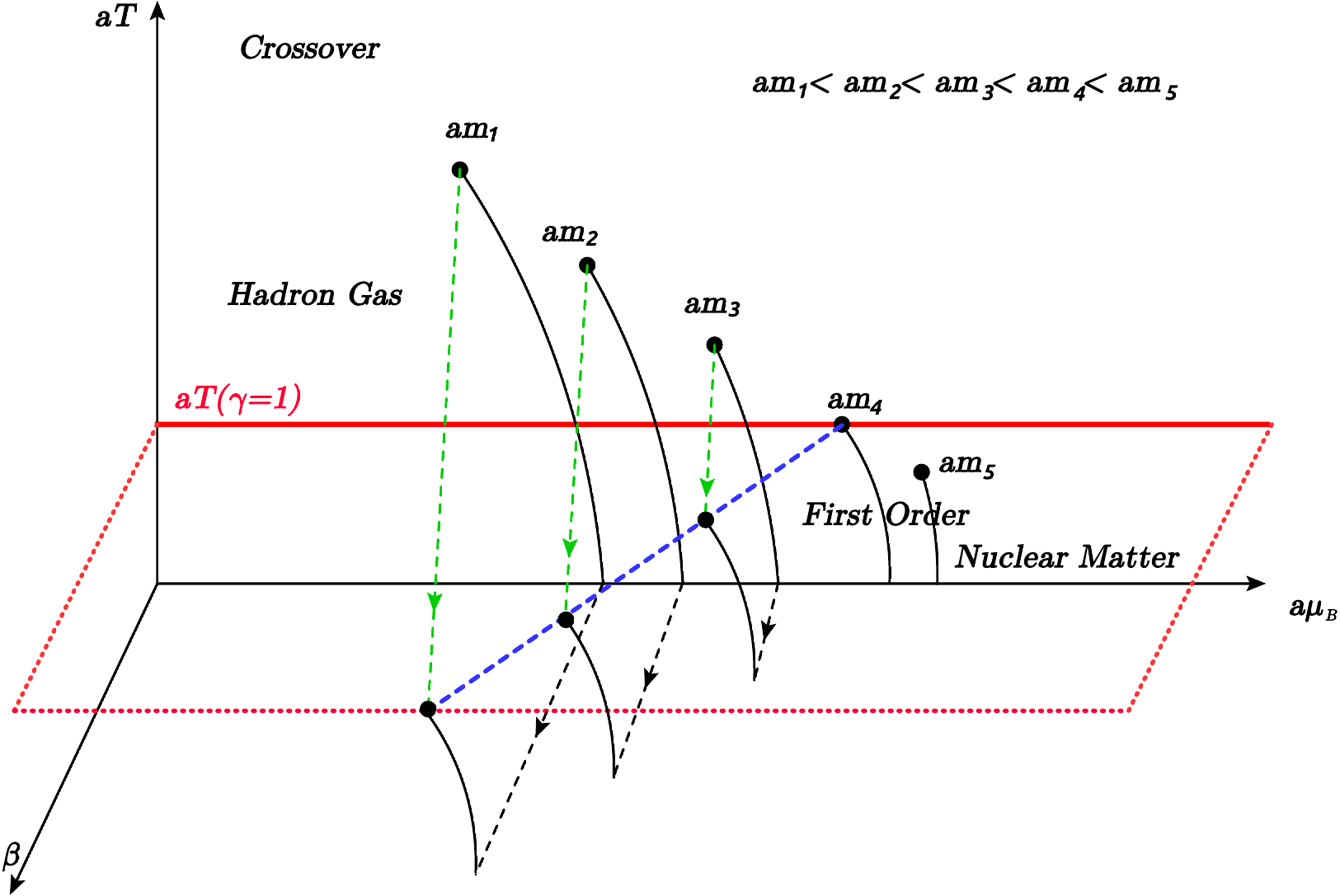}
\caption{\label{fig:diagram} 
Sketch of $\beta$ and quark mass dependence of the CEP on anisotropic lattices.
The red plane denotes fixed temperature in lattice units $aT=1/{N_t}$, and the
dotted blue line is the expected critical end line in the $aT=1/{N_t}$ plane. 
The dotted green lines are the expected behavior of CEP at fixed quark mass in
$aT>1/{N_t}$, as a function of $\beta$. 
}
\end{figure}

The finite density sign problem hinders the direct Monte Carlo simulation of QCD at
finite baryon chemical potential.
As an alternative method, we adopt the dual representation which is changing
degrees of freedom of the original theory to integer variables.
The dual representation in the strong coupling regime allows us to investigate the
full $\mu_B-T$ phase diagram.
We have studied the dependence of the nuclear critical end point (CEP) as a
function of the quark mass $am_q$ in the strong coupling limit
($\beta=0$)~\cite{Kim:2016izx}. 
If quark mass increases, the critical baryon chemical potential increases and
the critical temperature decreases.
Hence, as shown in Fig.~\ref{fig:diagram}, the critical end point moves to
the bottom right direction.
We also have presented the $\beta$ dependence in the chiral limit in our
previous study~\cite{Gagliardi:2017uag}.
In these proceedings, we present the $\beta$ dependence of the critical end line at
finite quark masses in the strong coupling regime.
We sketch the expected behavior of the critical line in Fig.~\ref{fig:diagram}. 
If $\beta$ increases, the critical end point of a certain quark mass is
expected to move to lower temperature but the critical baryon chemical
potential does not change much.
Hence, the first order line shortens with increasing $\beta$.

\section{Setup}
We use staggered fermions in the dual
formulation~\cite{Rossi:1984cv,Adams:2003cca,Fromm} with gauge corrections
$O(\beta)$ ~\cite{deForcrand:2014tha,Gagliardi:2017uag,Gagliardi:2019}.
\begin{align}
&Z(m_q,\mu,\gamma)=\nonumber\\
&\sum_{\{k,n,\ell,n_p\}}\underbrace{\prod_{b=(x,\mu)}\frac{(N_c-k_b)!}{N_c!(k_b-{|f_b|})!}{\gamma^{2k_b\delta_{\hat{0},\hat{\mu}}}} }_{\text{singlet hoppings}\, }
\underbrace{
  \prod_{x}\frac{N_c!}{n_x!}(2am_q)^{n_x}
}_{\text{chiral condensate}\,}
\underbrace{\prod_{\ell_3} w(\ell_3,\mu)}_{\text{triplet hoppings}\,}
\underbrace{\prod_{\ell_f} \tilde{w}(\ell_f,\mu)}_{\text{weight modification}}
\underbrace{\prod_P\frac{\left(\frac{\beta}{2N_c}\right)^{n_P+\bar{n}_P}}{n_P!\bar{n}_P!}}_{\text{gluon propagation}}\,, 
\end{align}
\begin{align}
w(\ell_3,\mu)=\frac{1}{\prod_{x \in \ell_3}}\sigma(\ell_3) \gamma^{N_c N_{\hat{0}}}\exp{(N_c N_t r_{\ell_3} a_t \mu)} \,, \quad
\sigma(\ell_3)=(-1)^{r_{\ell_3}+N_{-}(\ell_3)+1}\prod_{b=(x,\hat{\mu})\in \ell_3} \eta_{\hat{\mu}}(x)\,,
\end{align}
where $k_b$ and $f_b$ are the number of dimers and gauge fluxes at bond $b$,
$n_x$ is the number of monomers at site $x$, and $\ell_3$ denotes a 3-fermion
fluxes loop and $\ell_f$ is a single fermion loop.  $n_P$ and $\bar{n}_P$ are
plaquette (counterclockwise and clockwise) occupation number. 
$N_{\hat{0}}$ is the number of 3-fermion fluxes in temporal direction,
$r_{\ell_3}$ is the winding number in temporal direction.
$N_{-}$ is the number of 3-fermion fluxes in negative direction and
$\eta_{\hat{\mu}}$ is the staggered phase factor.
In this simulation, we fix the temporal lattice extent to $N_t=4$ and an anisotropy $\gamma=1$. The lattices temperature is fixed to $aT=\frac{1}{N_t}$.
If we change $\beta$, the lattice spacing $a$ is changed but $aT$ is invariant.
We simulate for $\beta=0.0,0.1,\cdots,0.9, 1.0$ and $am_q$ from $0.0$ to $0.5$
with step size $0.01$. 
We scan the baryon chemical potential $a\mu_B$ for each $(\beta,am_q)$ and find
the critical baryon chemical potential $a\mu_{c}$.
On the $\mu-T$ plane in Fig.\ref{fig:diagram}, if the quark mass becomes heavy,
   the critical end point moves to the large $a\mu$ and low $aT$ region.
Because we fix the temperature and vary the quark mass, the first order phase
transition occurs at lighter quark masses in our simulation.
By contrast, a crossover transition is expected for heavy quark mass.
At a certain critical quark mass ($am_4$ in Fig.\ref{fig:diagram}), the
transition turns into a second order transition.
\section{Analysis and results}
\subsection{Sign problem}
\begin{figure}
	\subfigure[From 3 fermion fluxes]{
		\label{fig:sign_c3}
\includegraphics[width=0.31\textwidth]{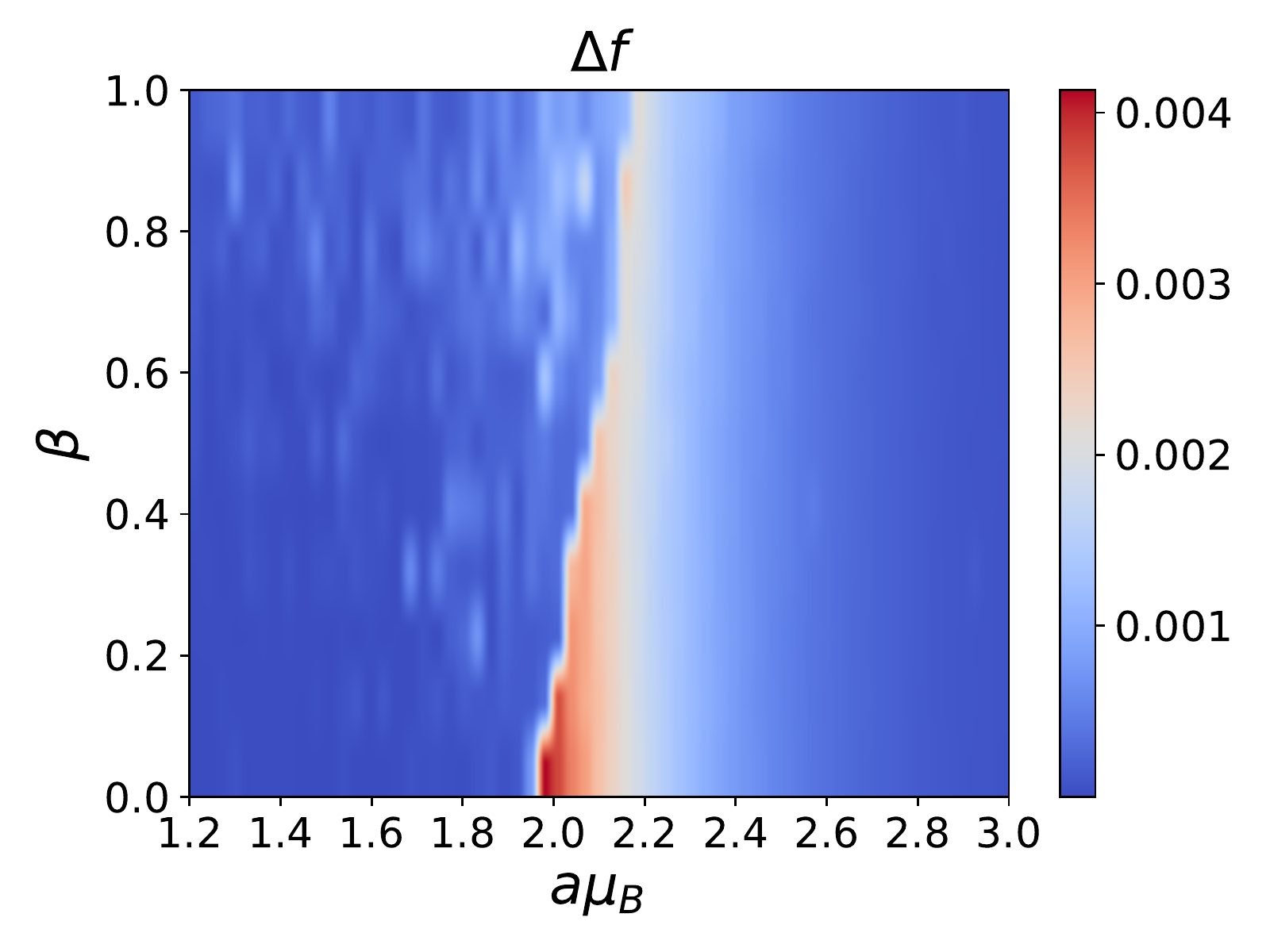}
}
	\subfigure[From Single fermion flux]{
		\label{fig:sign_c1}
\includegraphics[width=0.31\textwidth]{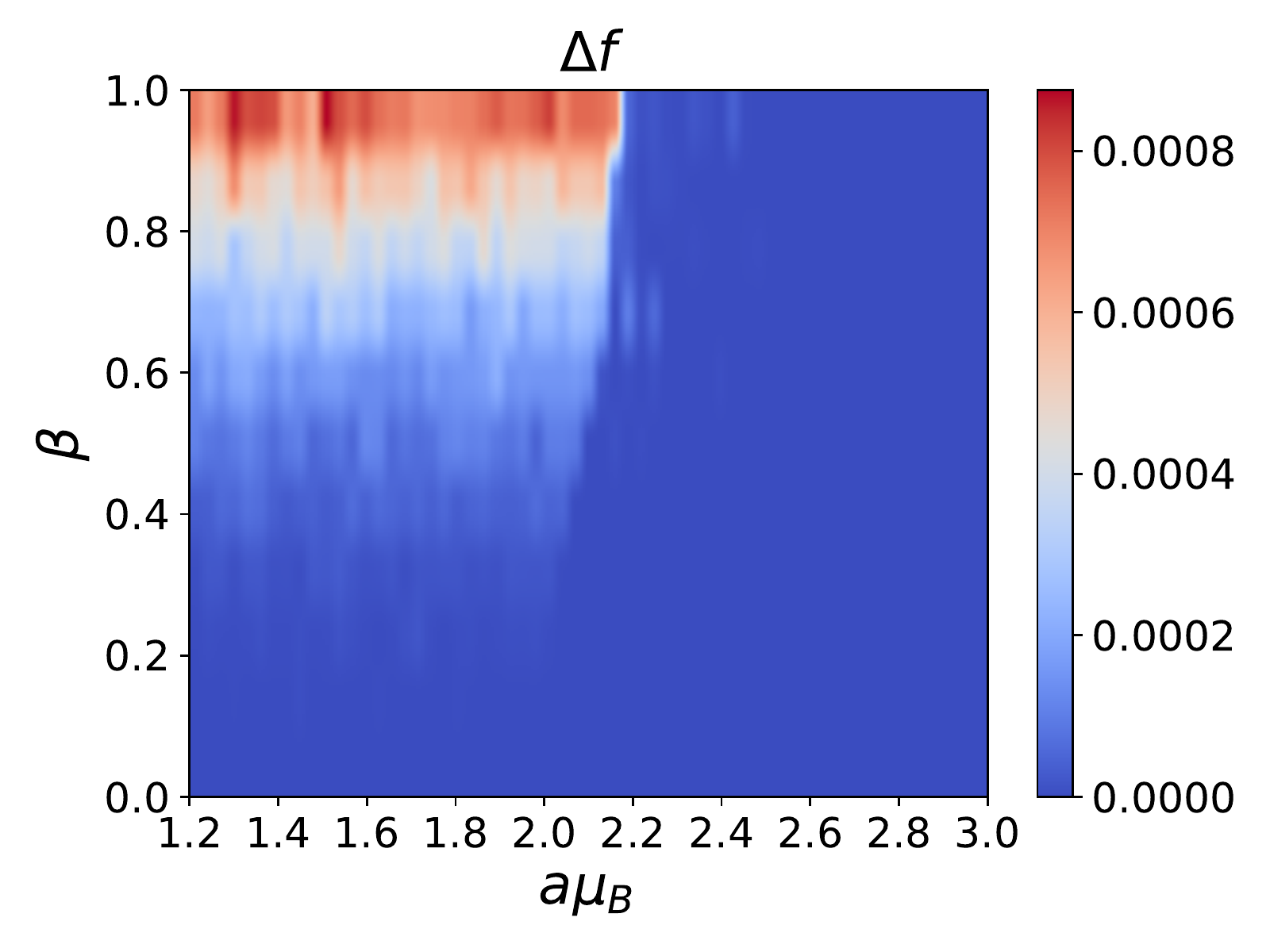}
}
	\subfigure[Total sign problem]{
		\label{fig:sign_tot}
\includegraphics[width=0.31\textwidth]{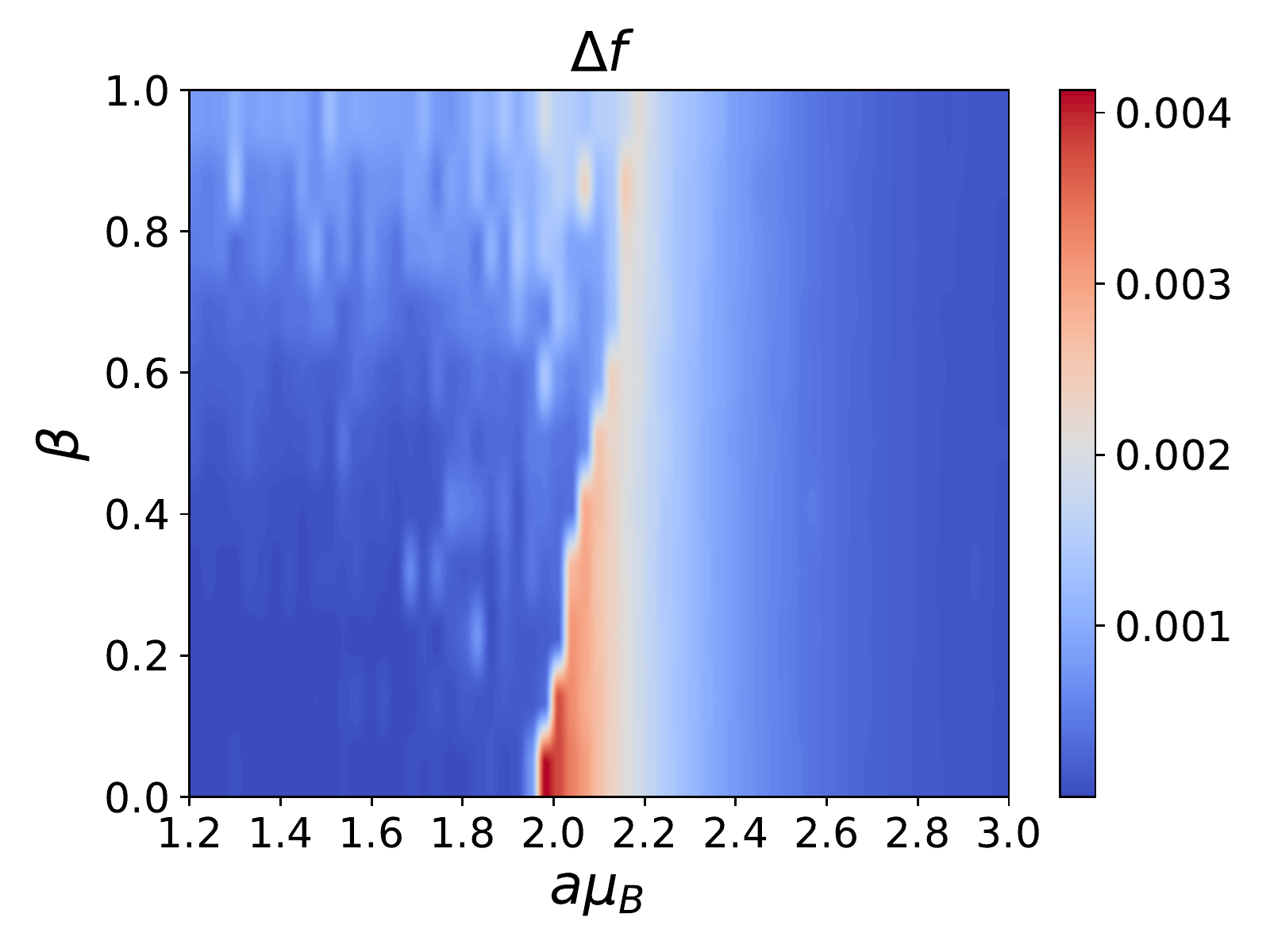}
}
\caption{\label{fig:sign} Sign problems from two difference sources.}
\end{figure}
First, let us consider the sign problem at finite $\beta$ and $am_q$.
The sign problem comes from the odd number of fermion flux. 
In $O(\beta)$, only single and 3 fermion fluxes cause the sign problem and they
are distinguishable.
Single fermion flux and 3 fermion fluxes make fermion loops like baryon world
lines in the strong coupling limit, and the sign problems is related to their
geometries~\cite{Gagliardi:2017uag}. 
We distinguish these two types of sign problems and show in the
Fig.~\ref{fig:sign_c3} and Fig.~\ref{fig:sign_c1}. 
Here $\Delta f$ is a difference between full and sign quenched free energy
density and sign is $\sigma=\exp{(-N_s^3 N_t \Delta f)}$.
In the Fig.~\ref{fig:sign_c3}, the sign problem from 3 fermion fluxes mainly
occurs near the phase transition region. 
We will see in the next section that the first order transition weakens when
increasing $\beta$. 
So, the phase transition area gets wide in $a\mu$.
The single fermion flux sign problem increases with $\beta$ as expected, but
only occurs in the hadronic gas phase as shown in Fig.~\ref{fig:sign_c1}.
We present the combined sign problem in Fig.~\ref{fig:sign_tot}.

\subsection{\label{sec:bar_dens} Analysis of baryon density}
\begin{figure}[t]
\subfigure[$\beta=0.1$, $am_q=0.3$]{
	\label{fig:bar11}
\includegraphics[width=0.48\textwidth]{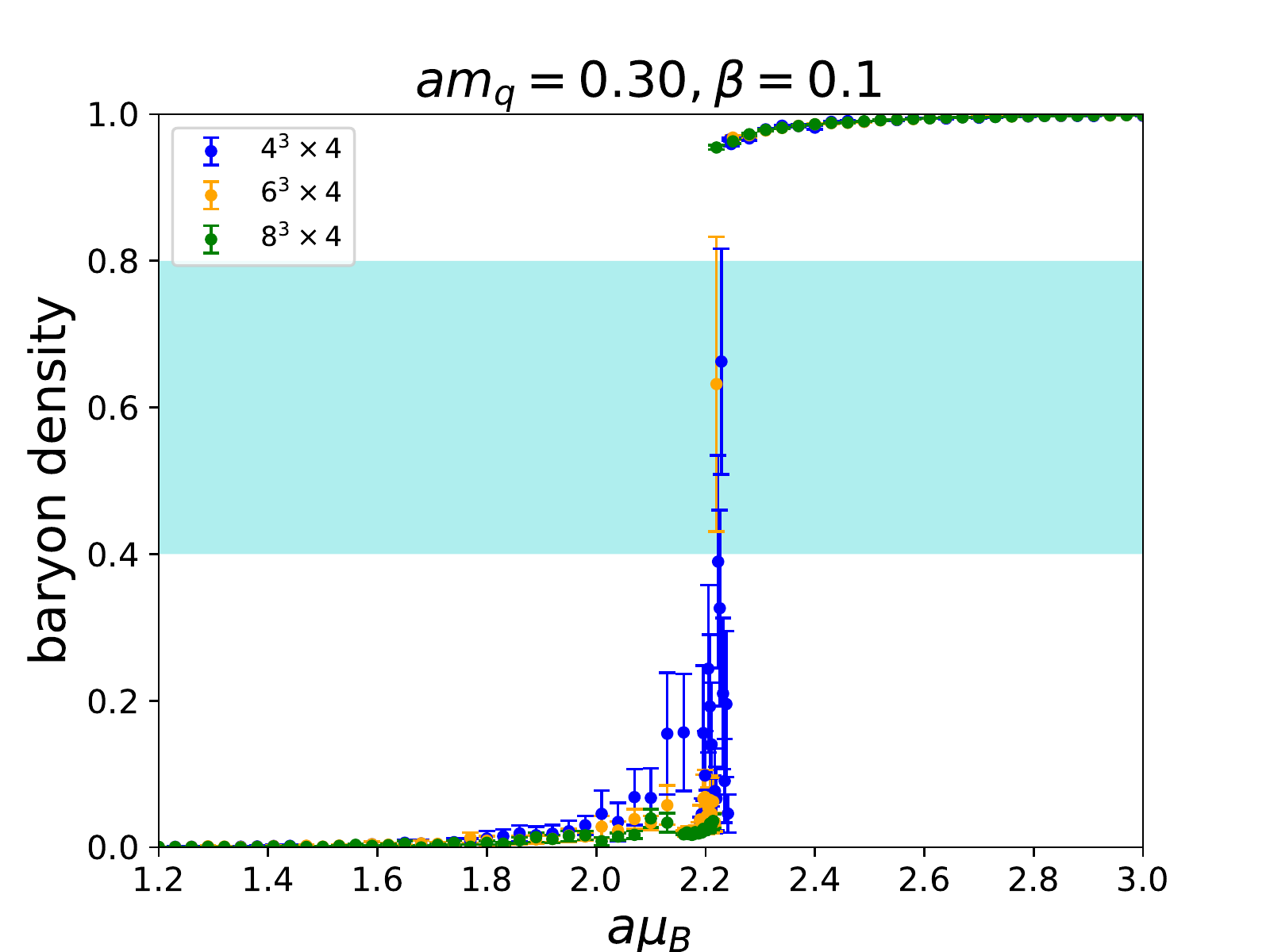}
}
	\subfigure[$\beta=0.1$, $am_q=0.95$]{
	\label{fig:bar12}
\includegraphics[width=0.48\textwidth]{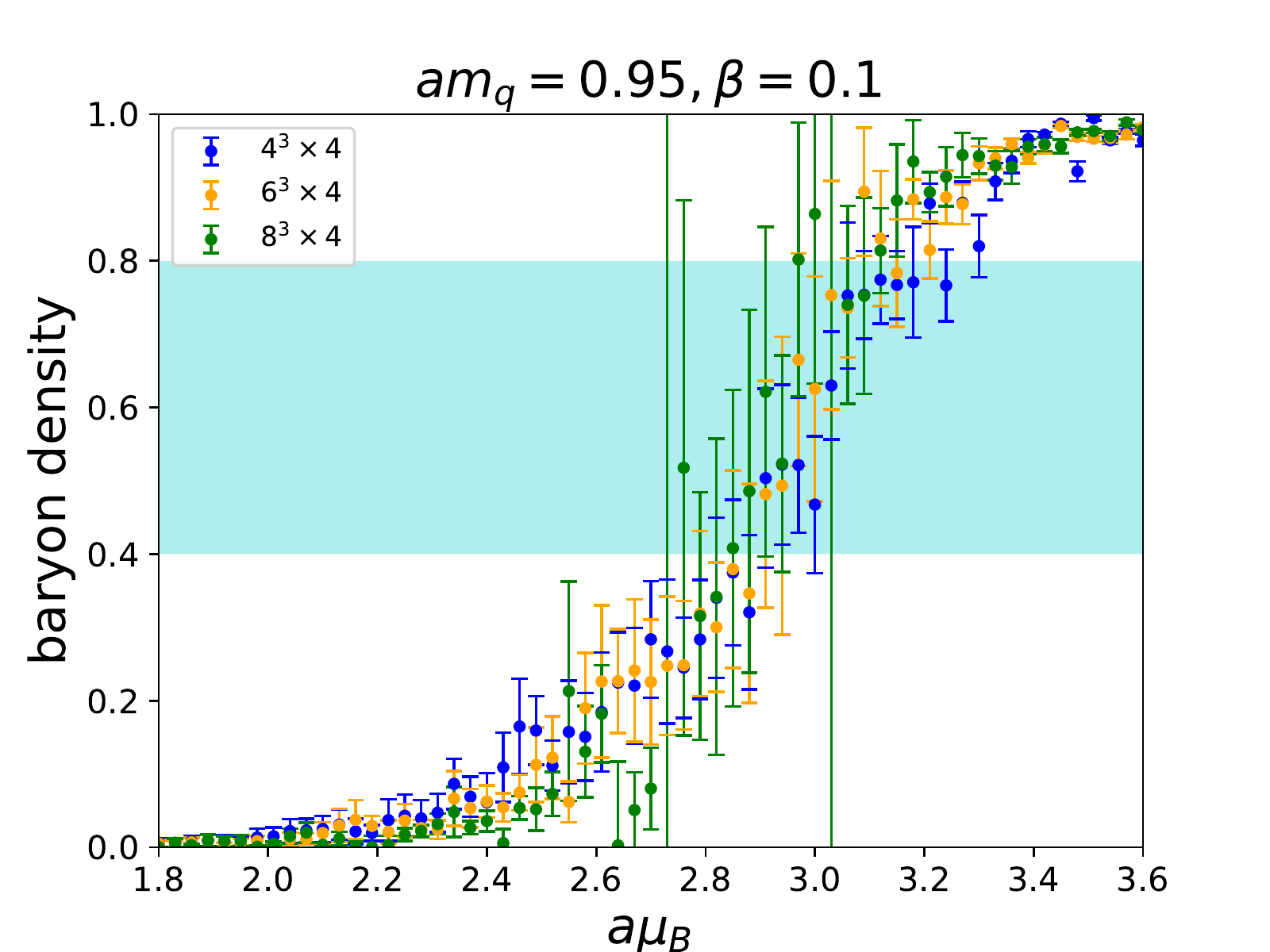}
}
	\caption{\label{fig:bar1} Baryon density at $\beta=0.1$}
\end{figure}

\begin{figure}[t]
	\subfigure[$\beta=1.0$, $am_q=0.3$]{
	\label{fig:bar21}
\includegraphics[width=0.48\textwidth]{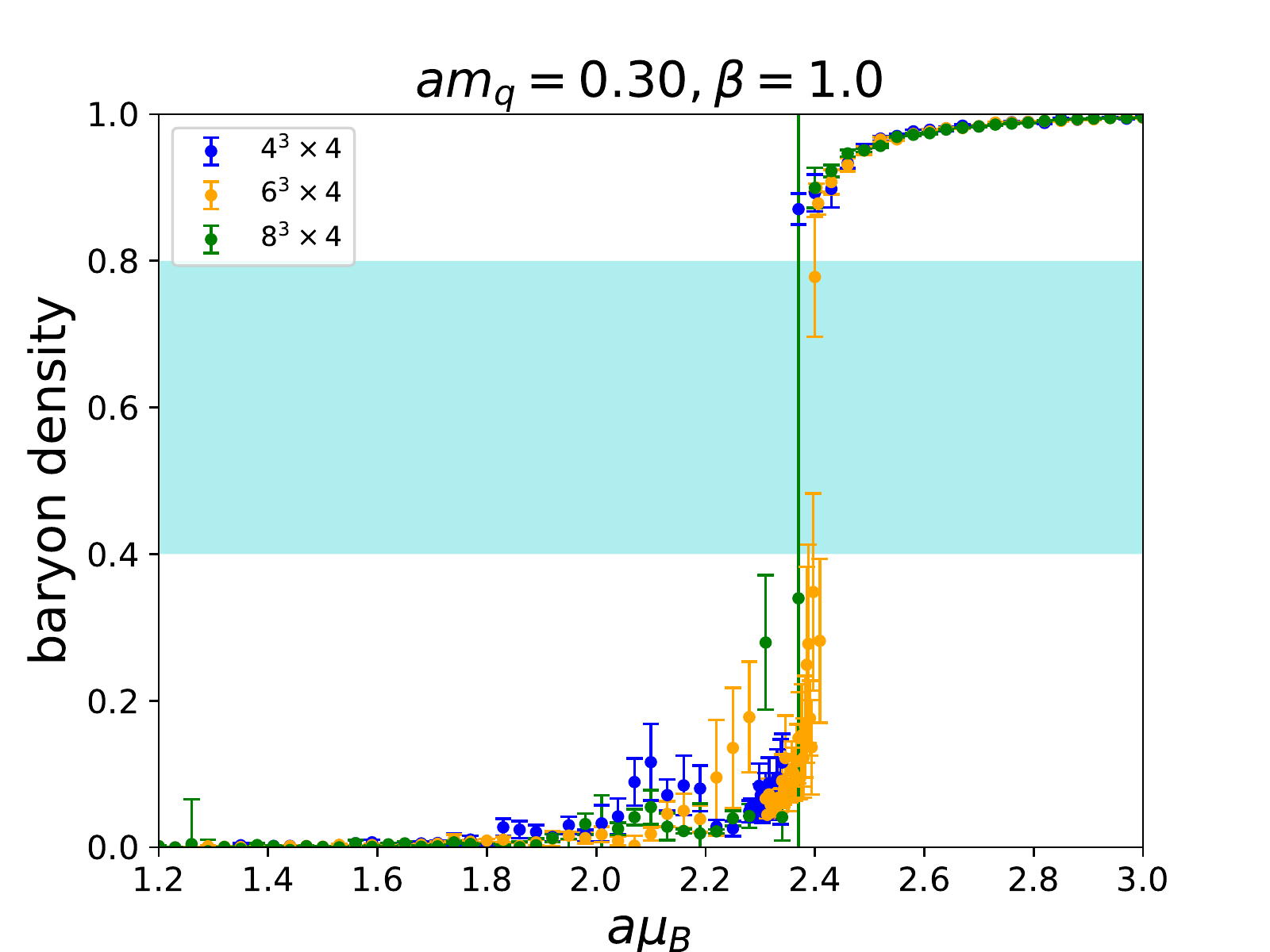}
}
	\subfigure[$\beta=1.0$, $am_q=0.95$]{
	\label{fig:bar22}
\includegraphics[width=0.48\textwidth]{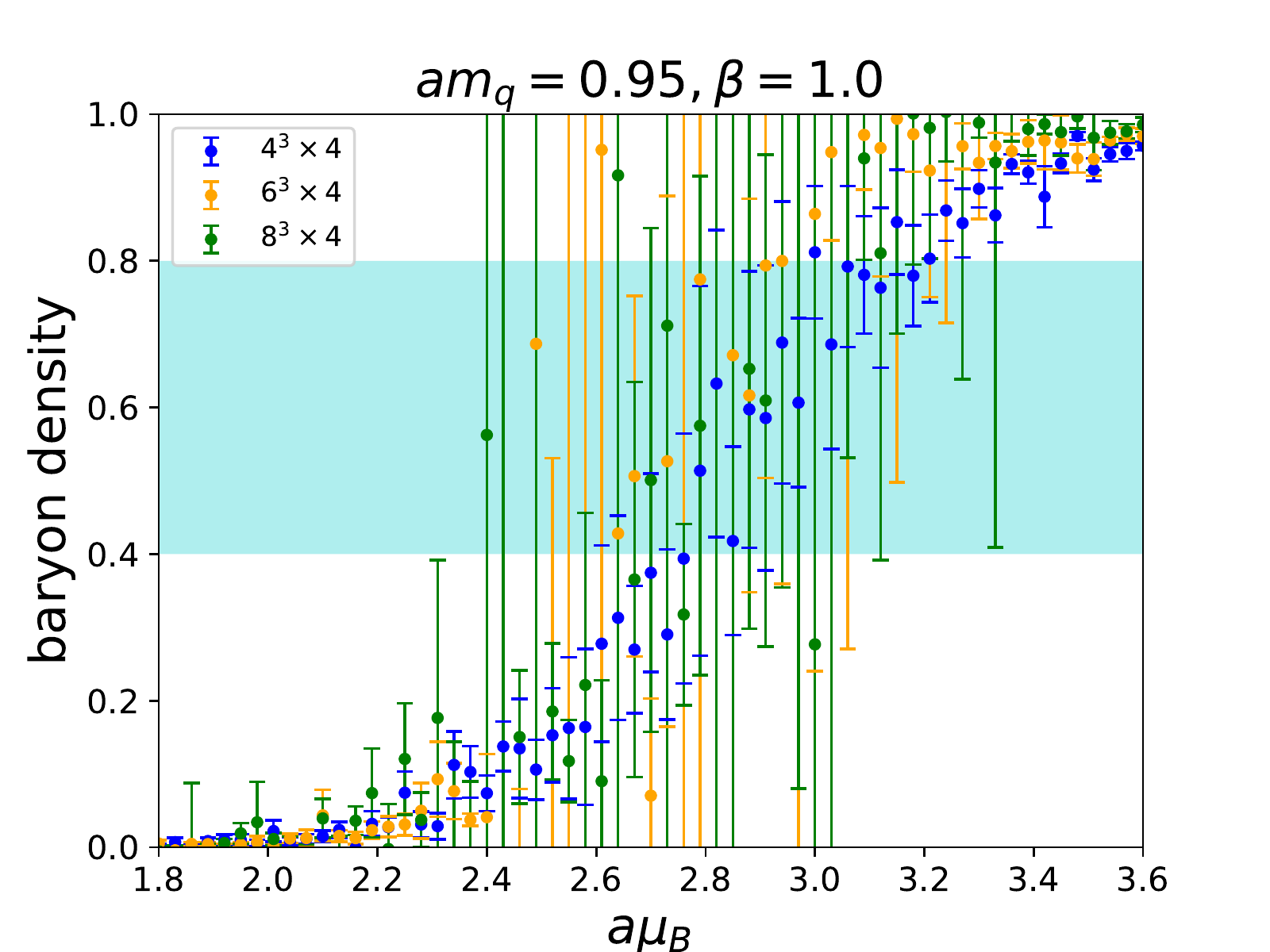}
}
	\caption{\label{fig:bar2} Baryon density at $\beta=1.0$}
\end{figure}
Now let us consider the baryon density $\langle n_B \rangle = \dfrac{1}{V_s N_t}\dfrac{\partial}{\partial(N_c a_t \mu)} \log Z =  \dfrac{1}{V_s} \langle \sum_{\ell_3} r_{\ell_3} \rangle$ which is our observable to determine the
critical end points, where $V_s$ is a spatial volume. 
We present our results of baryon density in Fig.~\ref{fig:bar1} and
Fig.~\ref{fig:bar2}.
In a first step, the critical $a\mu_c$ is determined by data points crossing the blue band, which was chosen by eye to be sufficiently distinct from both zero and saturation.
In Fig.~\ref{fig:bar1} we compare the onset of the nuclear transition for
$\beta=0.1$ for two different quark masses and various volumes. 
At small quark mass, the transition is consistent with first order,
and large quark mass weakens the phase transition. 
This also holds for Fig.~\ref{fig:bar2}.
If we fix the quark mass and change $\beta$, comparing Fig.~\ref{fig:bar11} and
Fig.~\ref{fig:bar21} or Fig.~\ref{fig:bar12} and Fig.~\ref{fig:bar22}, we can
see increasing $\beta$ diminishes the phase transition.
\begin{figure}
\subfigure[$\beta=0.0$]{
  \includegraphics[width=0.48\textwidth]{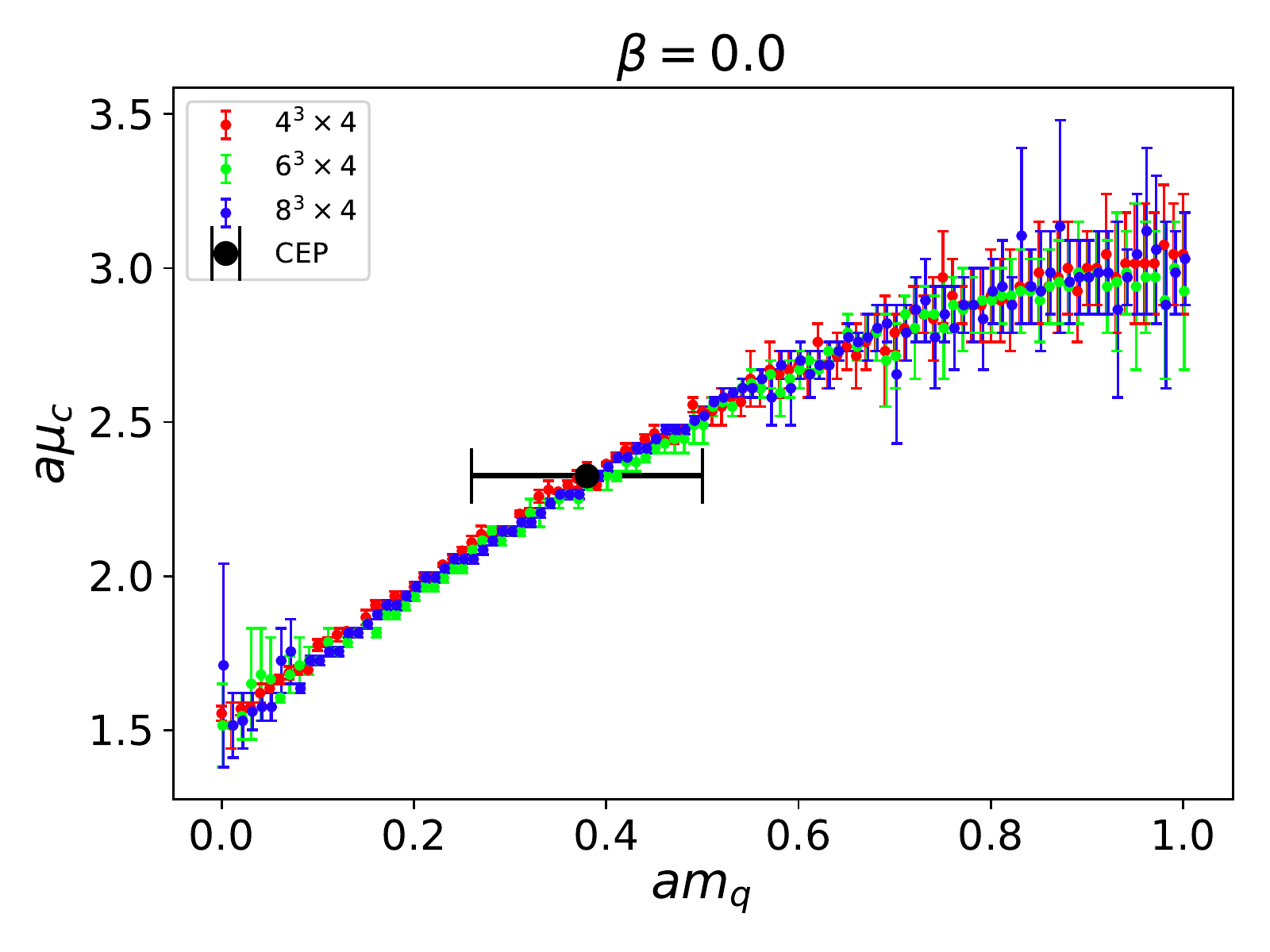}
}
\subfigure[$\beta=0.9$]{
  \includegraphics[width=0.48\textwidth]{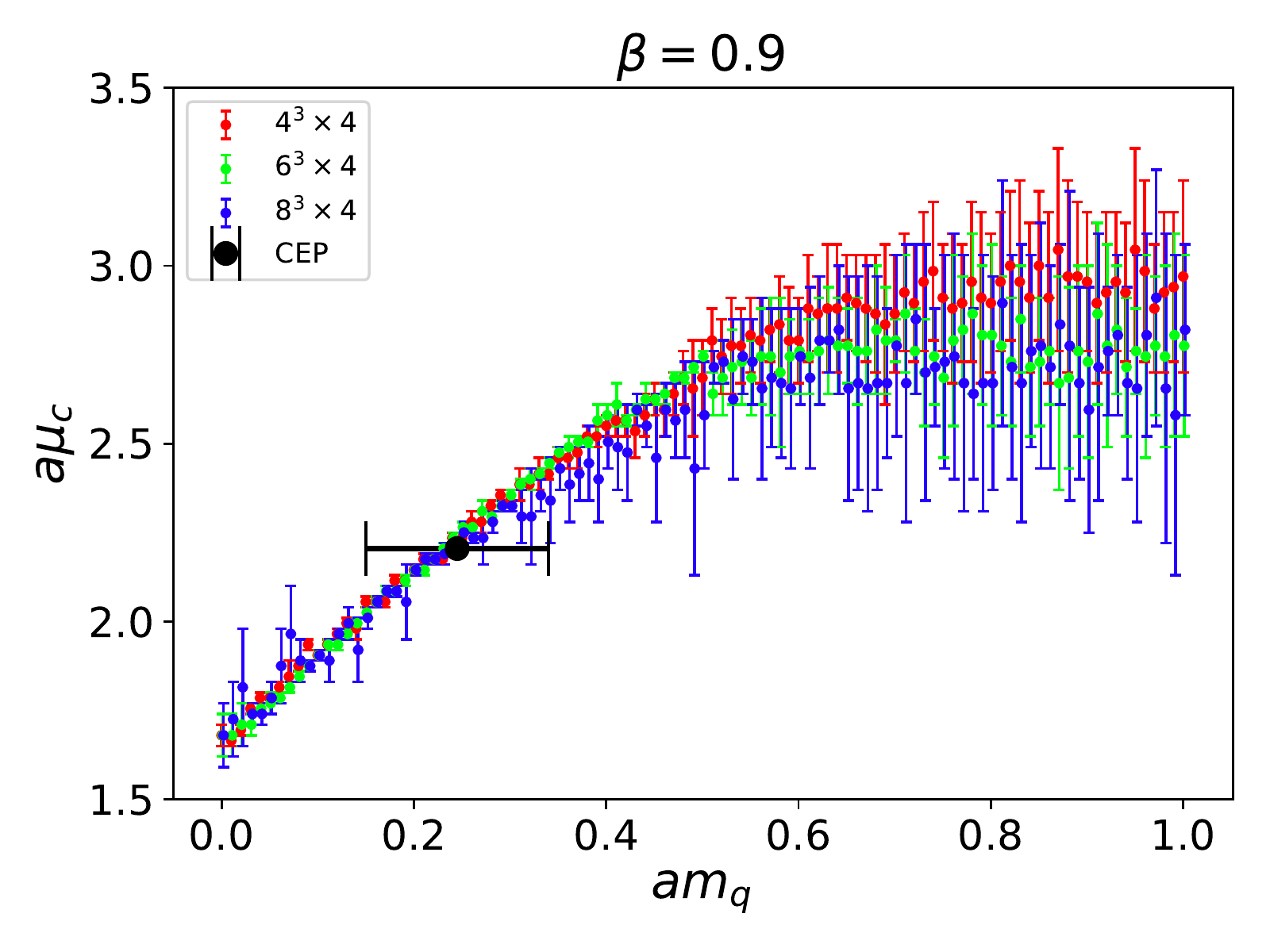}
}
\caption{\label{fig:amu_c} Critical chemical potential ($a\mu_c$) as a function of quark mass ($am_q$).}
\end{figure}
From the above analysis, we choose the $a\mu_c$ and plot them with respect to
$am_q$ in Fig.~\ref{fig:amu_c}.
The small quark mass region results in a first order phase transition and the
large quark region results in a crossover. 
As $\beta$ increases, the first order region shrinks and the crossover is extended. 
Between these first order and crossover transitions, there is a critical end
point. 
Hence, the critical end point is moving to the smaller quark masses when
$\beta$ is increased.
\subsection{Critical end point analysis using histograms}
\begin{figure}
\subfigure{
  \includegraphics[width=0.31\textwidth]{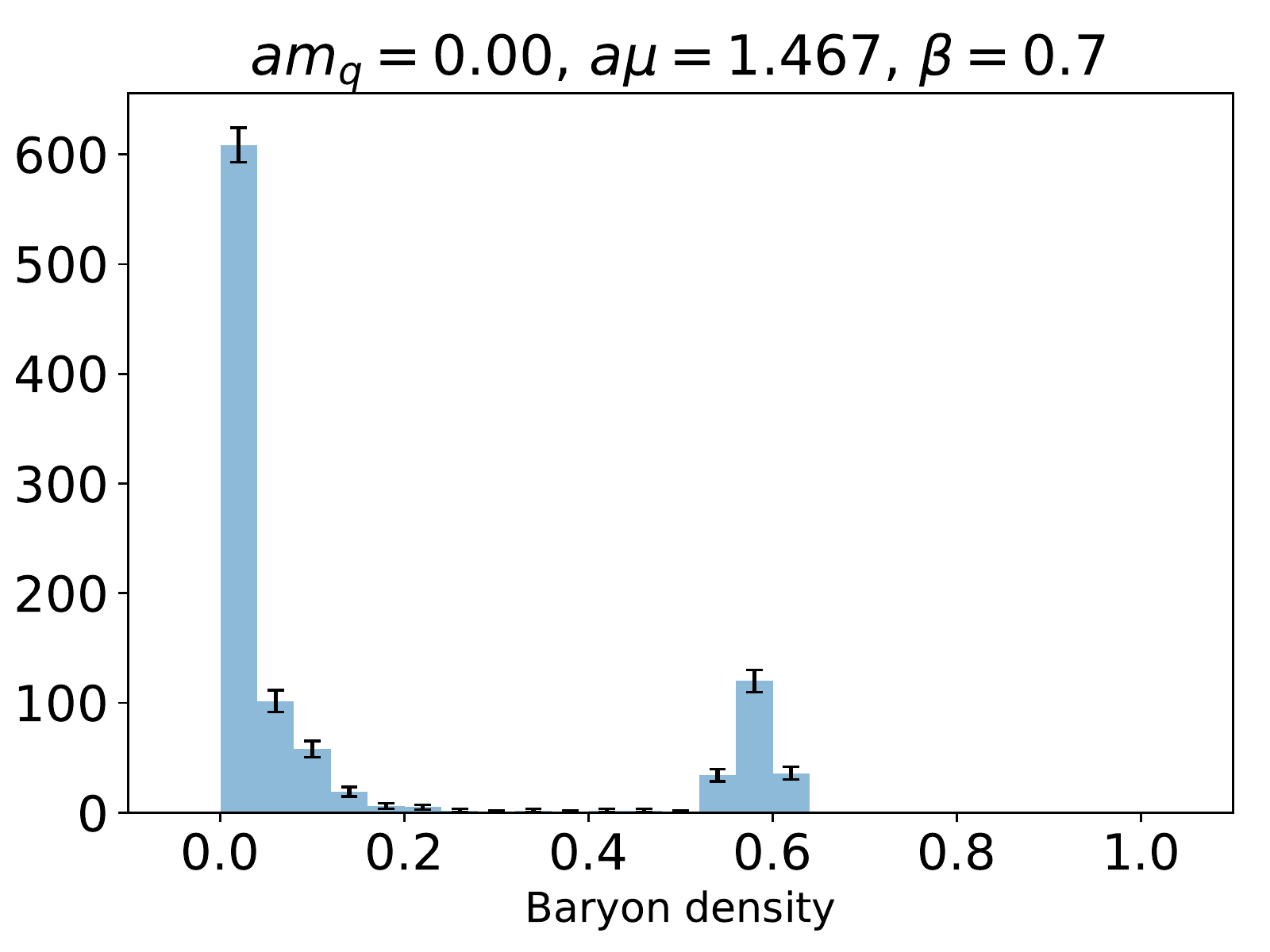}
}
\subfigure{
  \includegraphics[width=0.31\textwidth]{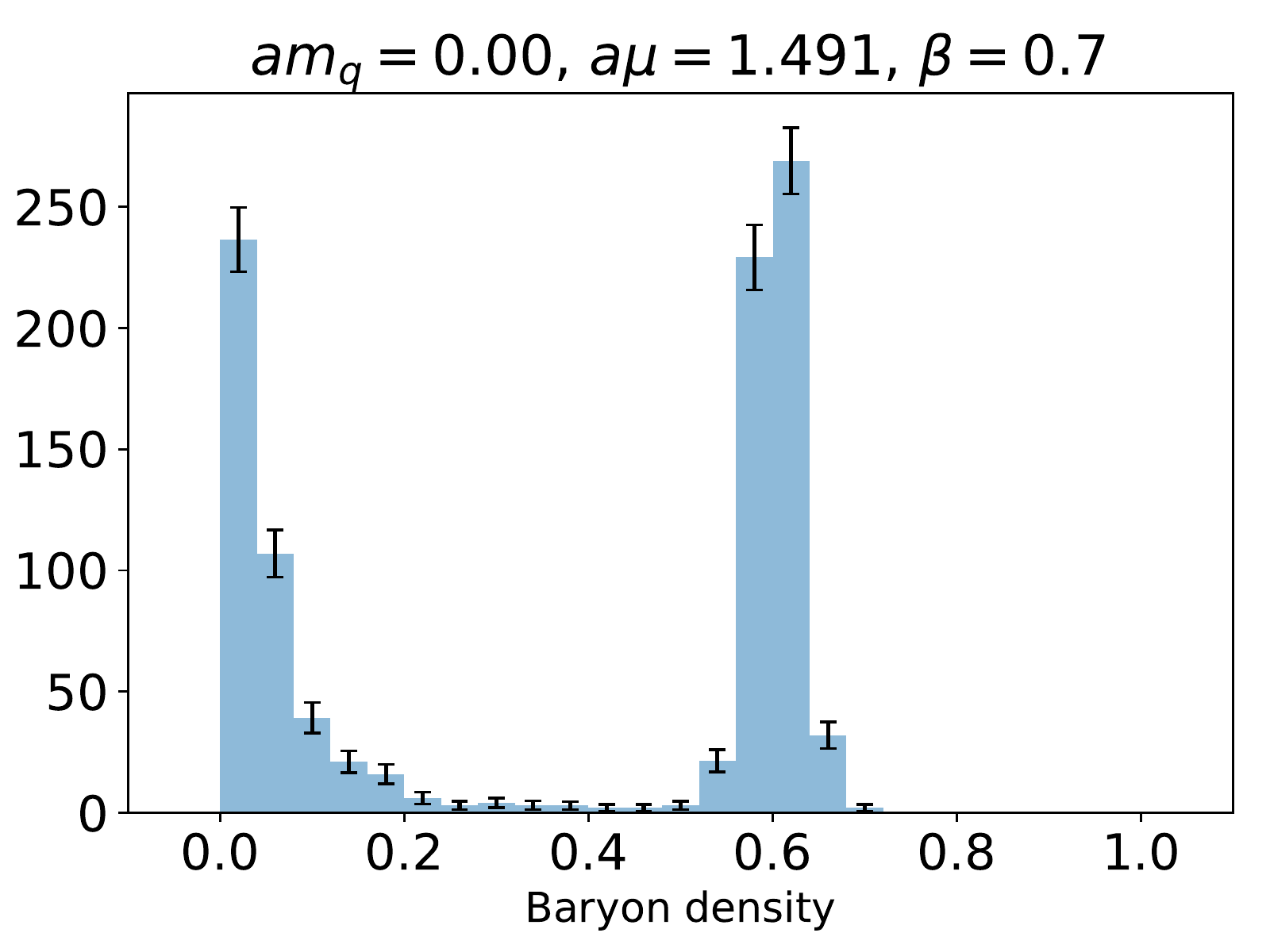}
}
\subfigure{
  \includegraphics[width=0.31\textwidth]{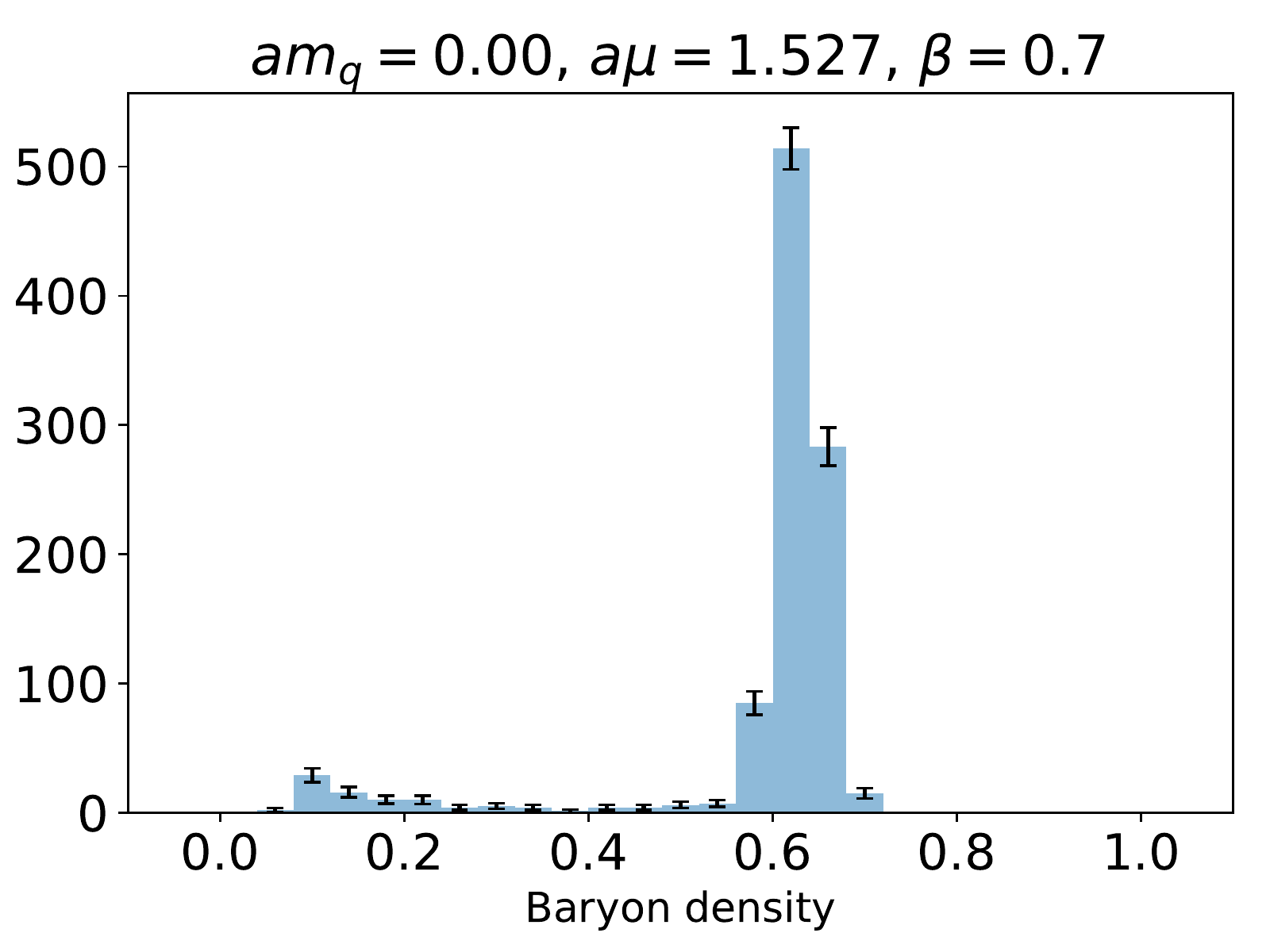}
}
\caption{\label{fig:histo_fitst} Baryon density histogram at small quark mass $am_q=0.0$ }
\end{figure}
\begin{figure}
\subfigure{
  \includegraphics[width=0.31\textwidth]{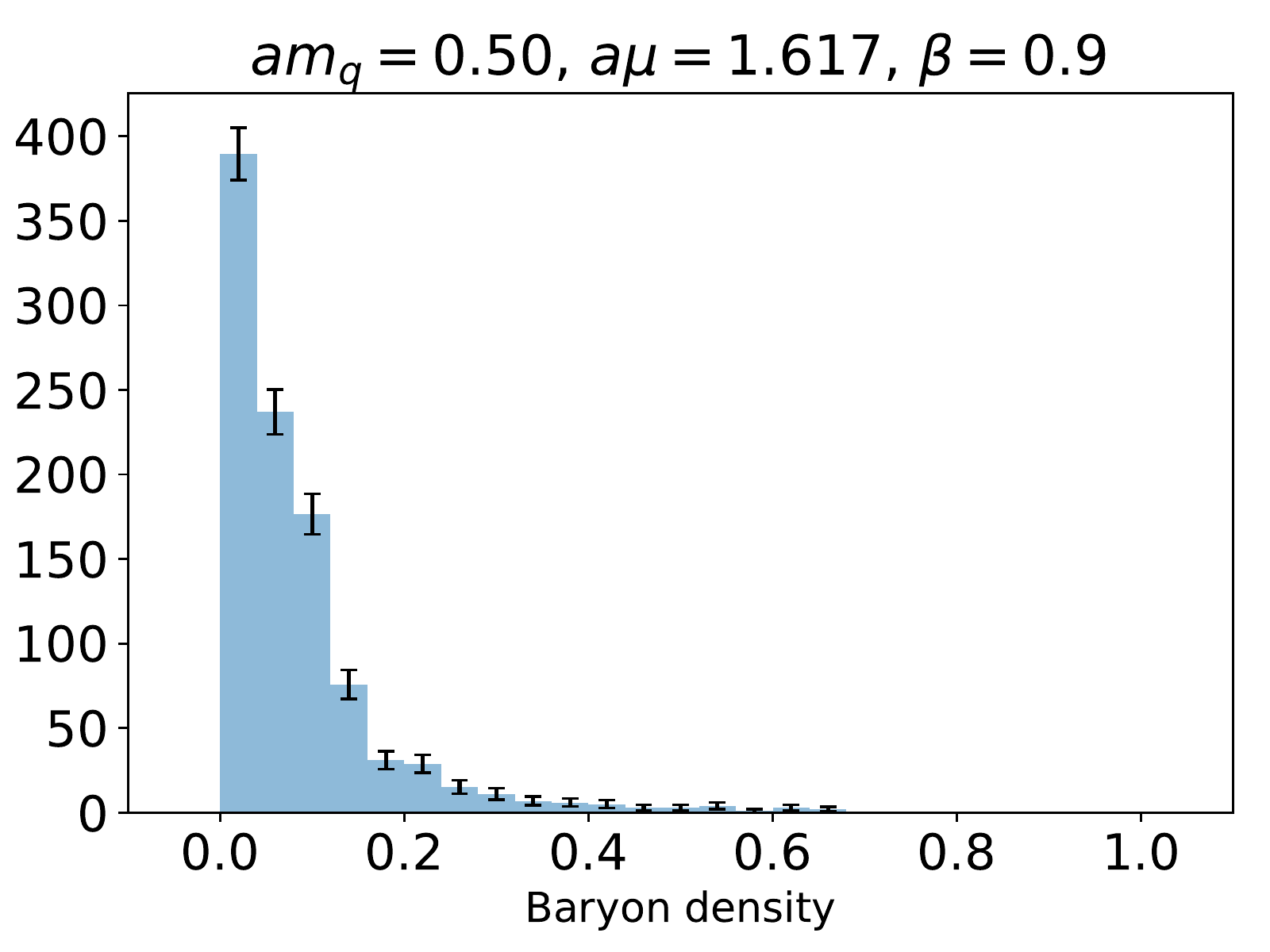}
}
\subfigure{
  \includegraphics[width=0.31\textwidth]{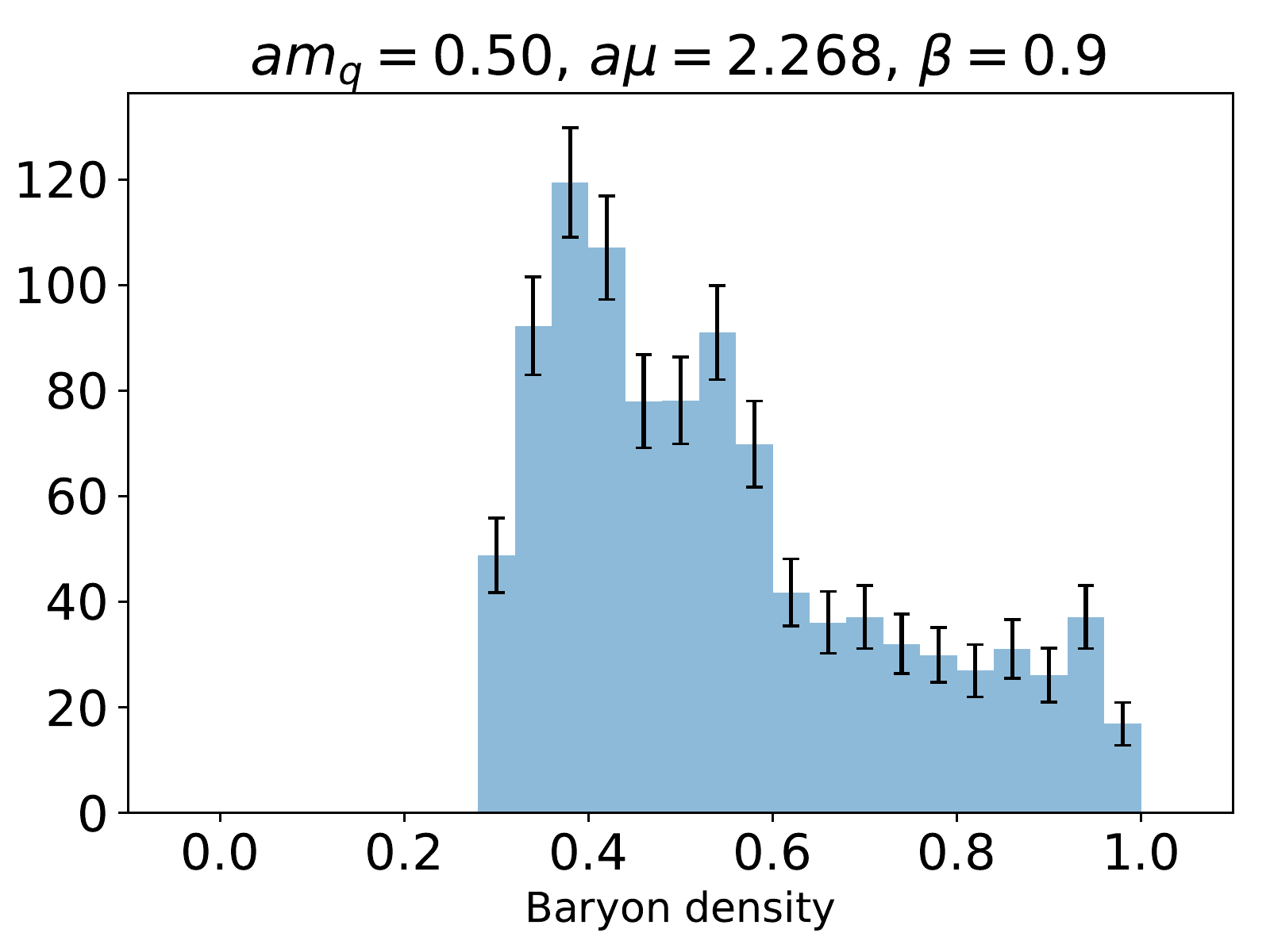}
}
\subfigure{
  \includegraphics[width=0.31\textwidth]{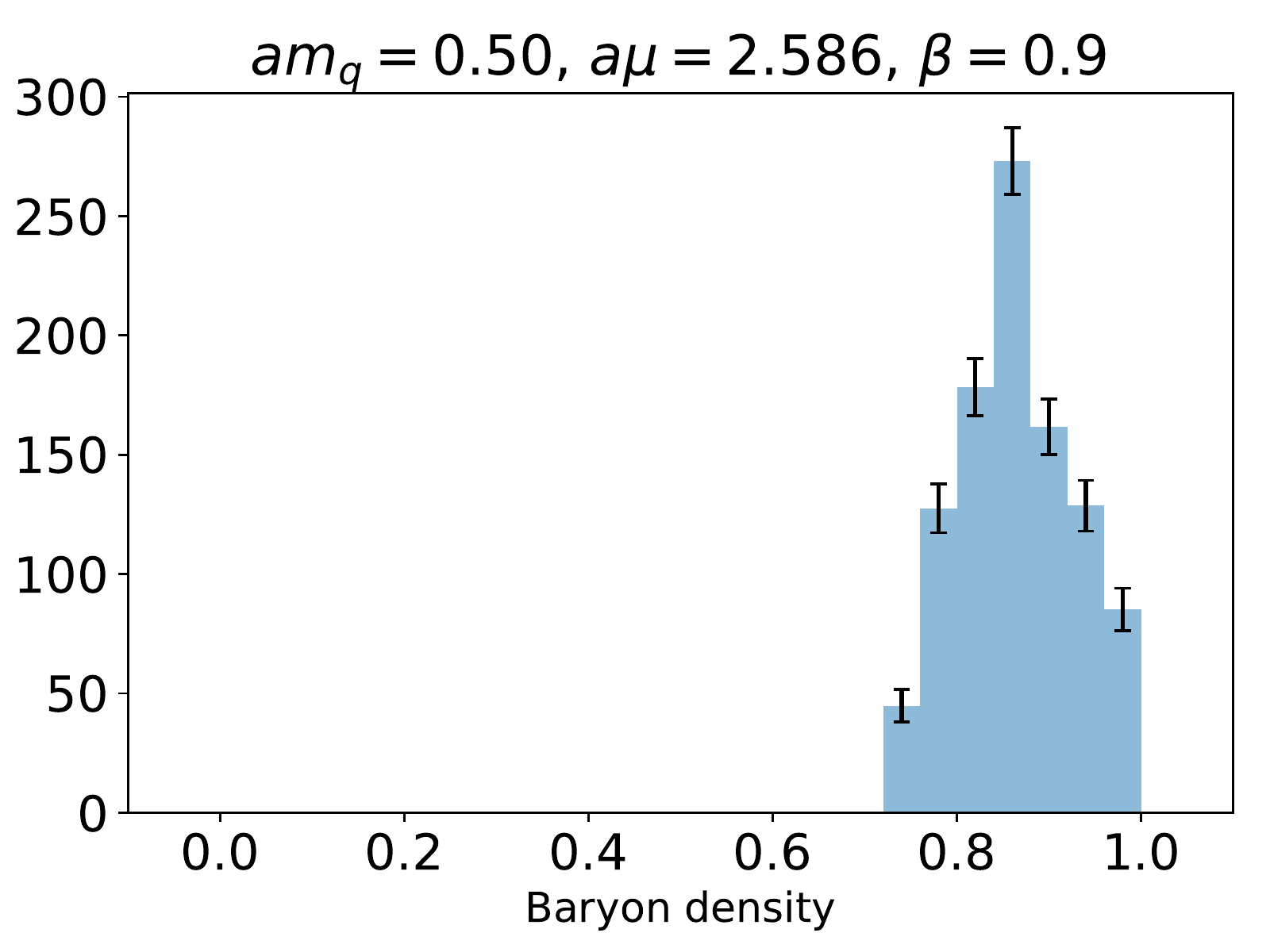}
}
\caption{\label{fig:histo_cross} Baryon density histogram at large quark mass $am_q=0.5$ }
\end{figure}
\begin{figure}[t]
\subfigure[Lower bound on CEP]{
  \label{fig:histo_two_peak}
  \includegraphics[width=0.48\textwidth]{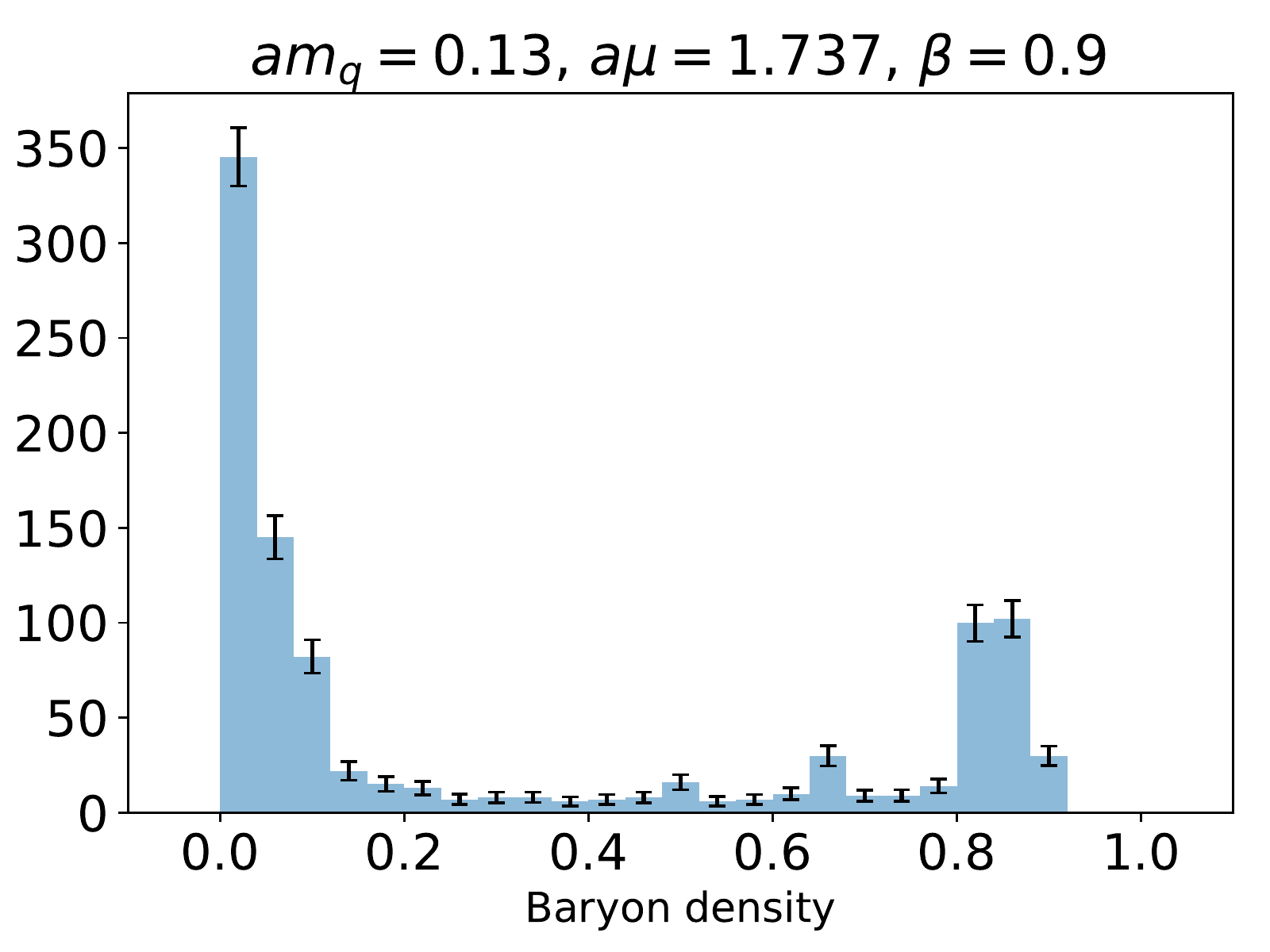}
}
\subfigure[Upper bound on CEP]{
  \label{fig:histo_no_peak}
  \includegraphics[width=0.48\textwidth]{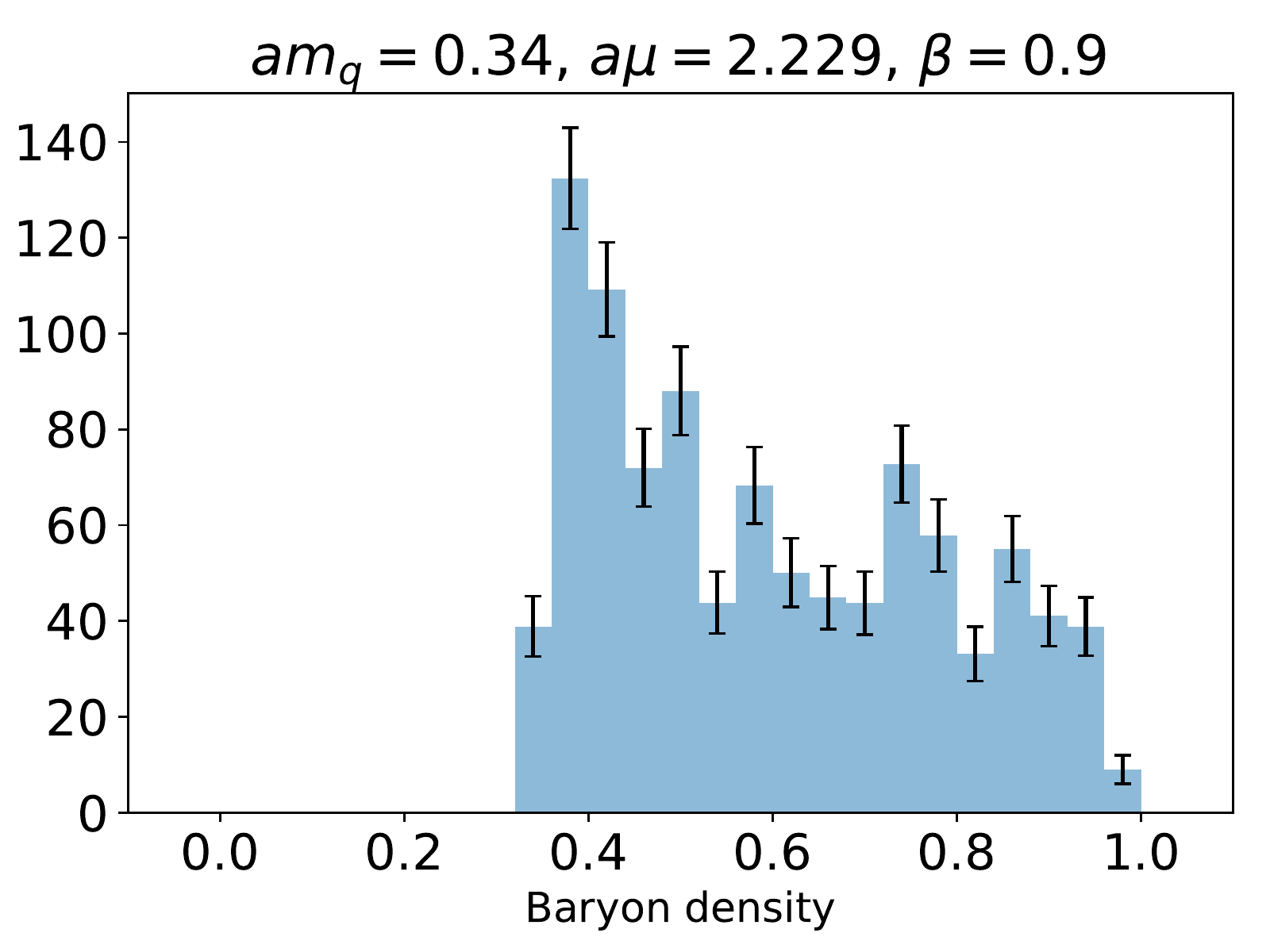}
}
\caption{\label{fig:histo_CEP} Baryon density histogram of lower and upper limit of $am_q$ at $a\mu_c$ for critical end point at $\beta=0.5$.}
\end{figure}
\begin{figure}
\subfigure[Quark mass critical line]{
  \label{fig:CEP_mq}
  \includegraphics[width=0.48\textwidth]{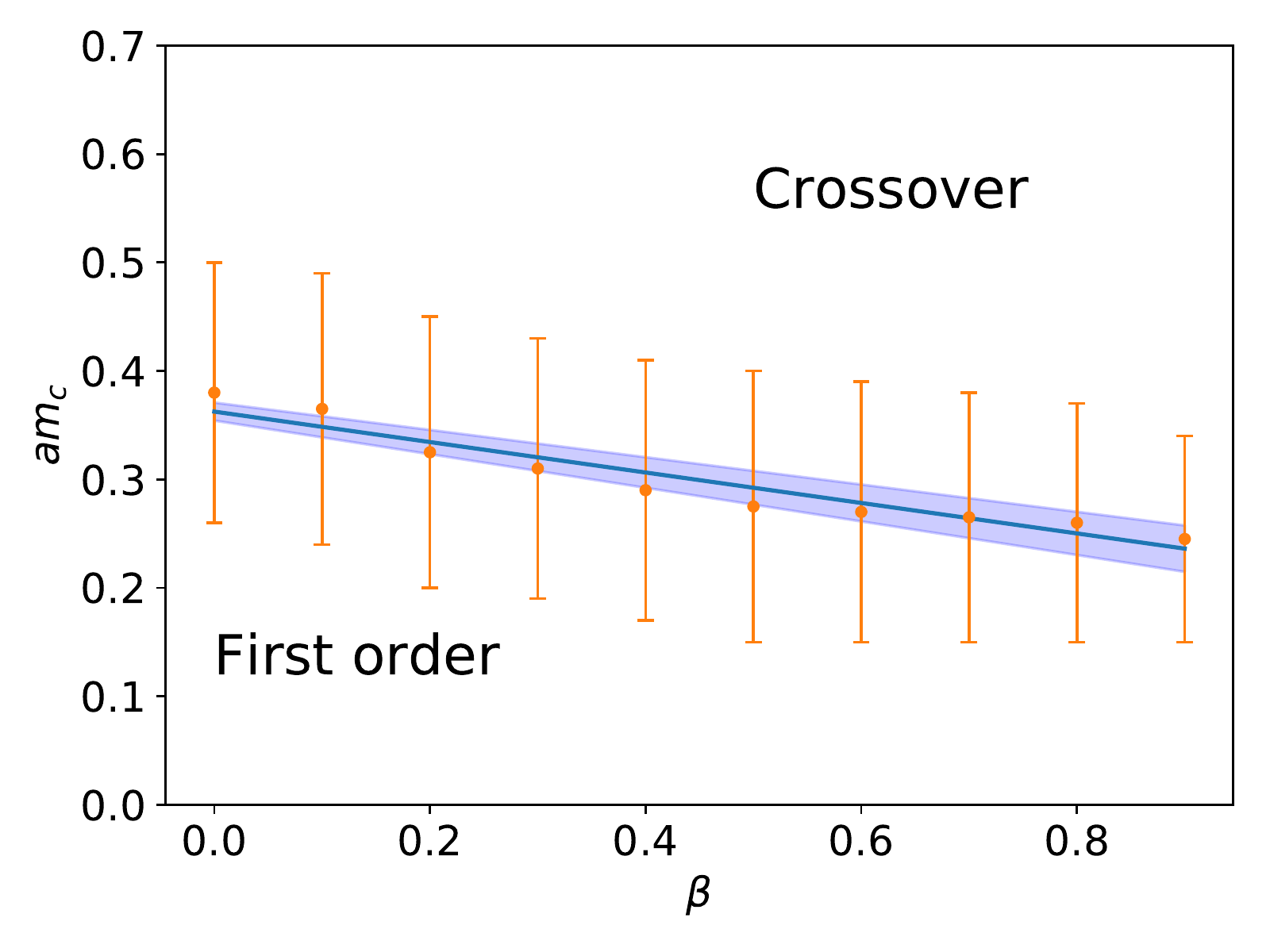}
}
\subfigure[Baryon chemical potential critical line]{
  \label{fig:CEP_mu}
  \includegraphics[width=0.48\textwidth]{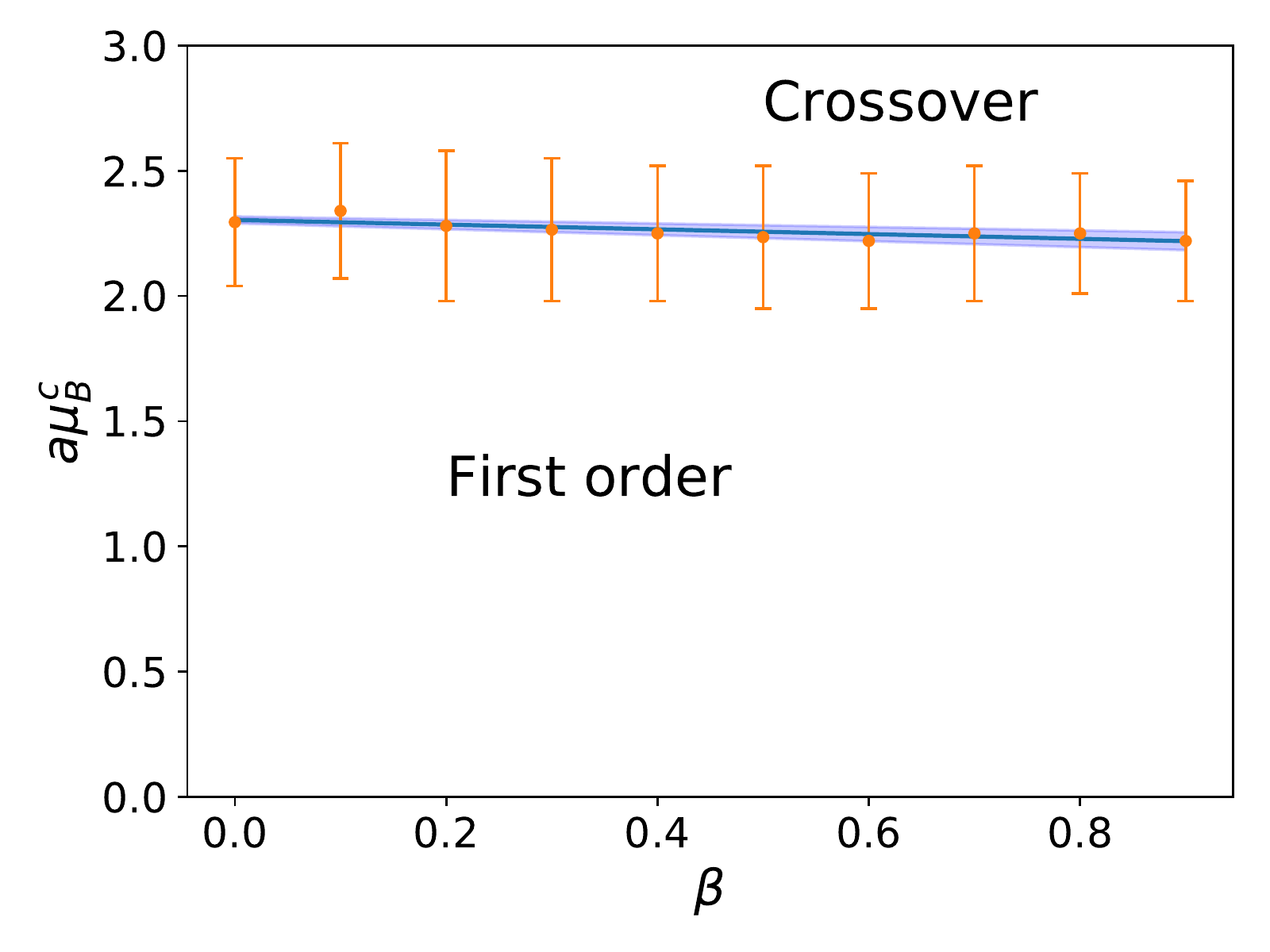}
}
\caption{\label{fig:endpoint} Critical line of end points as a function of $\beta$.}
\end{figure}

Because of the huge statistical errors near the critical chemical potential
$\mu_c$, the analysis of the baryon susceptibility is difficult. 
So, we analyse the data using histograms to bound the critical end points.
We use the data of a $8^3 \times 4$ volume for the histograms and the errors are
computed by the bootstrap resampling method.
In the Fig.~\ref{fig:histo_fitst} and Fig.~\ref{fig:histo_cross}, we plot
histograms around $a\mu_c$.
At small quark masses, the histograms have a two-peak distribution. 
Clearly at this quark mass, the phase transition is of first order.
For large quark mass, there is no two-peak distribution and the peak moves
smoothly from one state to the other state with increasing $a\mu$.
There is no perfect two-peak distributions in the first order transition
histogram because of finite volume effects, so we choose the lower bound for the
CEP when two-peak distribution is clear that is shown in
Fig.~\ref{fig:histo_two_peak}. 
The upper bound for the CEP is selected when two-peak vanishes and a peak starts
to move smoothly. 
This is presented in Fig.~\ref{fig:histo_no_peak}.
From the histogram analysis we can determine the upper limit of quark mass for
the first order phase transition and the lower limit of that for crossover.
Between those upper and lower limits, the second order end point is located.
The critical line as a function of $\beta$ is presented in
Fig.~\ref{fig:CEP_mq}. 
In this plot, the left lower corner corresponds to the first order region and
the right upper corner is crossover. 
As $\beta$ increases, the quark mass of the CEP decreases slightly. 
This is similar for the baryon chemical potential as shown in
Fig.~\ref{fig:CEP_mu}. 
To determine the error bars in Fig.~\ref{fig:CEP_mu}, we take the smallest and
largest quark masses in Fig.~\ref{fig:CEP_mq} for each $\beta$. 
Then there are corresponding $a\mu_c$ for quark masses. We use the range of
these $a\mu_c$ values to the errors of critical line in Fig.~\ref{fig:CEP_mu}.
We determine the error bars from the histograms very conservatively.
The huge errors in Fig.~\ref{fig:endpoint} are caused by the uncertainty and
small statistics. 
The $\beta$ dependence of both $am_q$ and $a\mu$ are linear in $\beta$ in the
small $\beta \lesssim 1$ region.

\section{Conclusion}
We simulate the dual representation with finite quark mass and lattice gauge
coupling $\beta$. 
Because this simulation takes into account only $O(\beta)$, we restrict to the
range of $\beta$ smaller than one. 
In this parameter space, the sign problem is still mild enough to use a sign
reweighting method. 
We obtain the $\beta$ dependence of the critical line using a histogram analysis.
In the small $\beta$ region, the critical line looks like linear, but still has large errors.
Hence, the higher corrections of $\beta$ are essential to extend this study to
$\beta$ larger than one and an important step in this direction has been
addressed in Ref.~\cite{Gagliardi:2019}.

\acknowledgments
The authors J.K., O.P. and W.U. acknowledge support by the Deutsche
Forschungsgemeinschaft (DFG, German Research Foundation) through the
CRC-TR 211 'Strong-interaction matter under extreme conditions'– project number
315477589 – TRR 211.  
J.K. would like to thank the Center for Scientific Computing, University of
Frankfurt for making their High Performance Computing facilities available.
W.U. acknowledges support by the Deutsche Forschungsgemeinschaft (DFG)
  through the Emmy Noether Program under Grant No.  UN 370/1.  

\bibliography{refs}

\providecommand{\href}[2]{#2}\begingroup\raggedright\begin{thebibliography}{1}

\bibitem{Kim:2016izx}
J.~Kim and W.~Unger, {\it {Quark Mass Dependence of the QCD Critical End Point
  in the Strong Coupling Limit}},  {\em PoS} {\bf LATTICE2016} (2016) 035,
  [\href{http://xxx.lanl.gov/abs/1611.09120}{{\tt 1611.09120}}].

\bibitem{Gagliardi:2017uag}
G.~Gagliardi, J.~Kim, and W.~Unger, {\it {Dual Formulation and Phase Diagram of
  Lattice QCD in the Strong Coupling Regime}},  {\em EPJ Web Conf.} {\bf 175}
  (2018) 07047, [\href{http://xxx.lanl.gov/abs/1710.07564}{{\tt 1710.07564}}].

\bibitem{Rossi:1984cv}
P.~Rossi and U.~Wolff, {\it {Lattice {QCD} With Fermions at Strong Coupling: A
  Dimer System}},  {\em Nucl. Phys.} {\bf B248} (1984) 105--122.

\bibitem{Adams:2003cca}
D.~H. Adams and S.~Chandrasekharan, {\it {Chiral limit of strongly coupled
  lattice gauge theories}},  {\em Nucl. Phys.} {\bf B662} (2003) 220--246,
  [\href{http://xxx.lanl.gov/abs/hep-lat/0303003}{{\tt hep-lat/0303003}}].

\bibitem{Fromm}
M.~Fromm, {\it {Lattice QCD at string coupling: thermodynamics and nuclear
  physics}},  {\em Thesis} (2010).

\bibitem{deForcrand:2014tha}
P.~de~Forcrand, J.~Langelage, O.~Philipsen, and W.~Unger, {\it {Lattice QCD
  Phase Diagram In and Away from the Strong Coupling Limit}},  {\em Phys. Rev.
  Lett.} {\bf 113} (2014), no.~15 152002,
  [\href{http://xxx.lanl.gov/abs/1406.4397}{{\tt 1406.4397}}].

\bibitem{Gagliardi:2019}
G.~Gagliardi and W.~Unger, {\it {A new dual representation for staggered
  lattice QCD}},  \href{http://xxx.lanl.gov/abs/1911.08389}{{\tt 1911.08389}}.

\end{thebibliography}\endgroup

\end{document}